\documentclass{optica-article}

\journal{opticajournal} 

\articletype{Research Article}

\usepackage{lineno}
\nolinenumbers

\begin{document}

\title{High-Order Exceptional Point-Based Rotation Sensing in Anti-Parity time Symmetric Microresonators}

\author{Wenxiu Li,\authormark{1} Zelei Li,\authormark{2}Jincheng Li,\authormark{2} Xin Sun,\authormark{1}  Zongqi Yang,\authormark{1} Ce Qin,\authormark{2} Xinyao Huang, \authormark{2} Anping Huang,\authormark{1} Hao Zhang \authormark{3,*} and Zhisong Xiao\authormark{2}}

\address{\authormark{1}School of Electronic Information Engineering, Beihang University, Beijing 100191, China
\authormark{2}School of Physics, Beihang University, Beijing 100191, China \\
\authormark{3}SchooL of Space and Earth Sciences, Beihang University, Beijing 100191, China}

\email{\authormark{*}haozhang@buaa.edu.cn} 


\begin{abstract*} 
Exceptional points (EPs), which arise from non-Hermitian systems, have been extensively investigated for the development of  high-performance gyroscopes. However, operating a non-Hermitian gyroscope at high-order EP (HOEP) to achieve extreme performance requires strict and precise control of parameters. Here, we propose the design of an anti-parity-time (anti-PT) symmetric optical gyroscope operating at a fourth-order EP, achieving both ultra-sensitivity and high robustness. Our configuration exhibits eigenfrequency splitting two orders of magnitude higher than that of anti-PT gyroscopes operating at second-order EP. Furthermore, we demonstrate a significant reduction in angular random walk (ARW) under noise limits, compared to anti-parity symmetric gyroscopes based on second-order EP. Our results provide a novel approach for developing high-sensitivity rotation detection based on HOEPs.

\end{abstract*}

\section{Introduction}
 Resonant-micro optical gyroscopes, based on the Sagnac effect, face a trade-off between compact size and high sensitivity\cite{01-khial2018nanophotonic,02-li2017microresonator,03-liang2017resonant,04-zhang2016chip}. With the miniaturization and integration of resonant optical gyroscopes, the reduced size of the resonant cavity loop typically results in a significant decrease in the Sagnac scaling factor. Recently, novel non-Hermitian optical gyroscopes operating near an EP have demonstrated superior performance in rotation detection\cite{05-hokmabadi2019non,06-lai2019observation,07-jiang2019chip,08-ren2017ultrasensitive,09-de2022design,10-de2019high,11-wang2024novel}. Exceptional points arise from non-Hermitian photonic systems, where eigenvalues degenerate and the corresponding eigenmodes coalesce \cite{12-chang2014parity,13-el2018non,14-guo2009observation,15-hodaei2014parity,16-jin2018parity,17-peng2014parity,18-ruter2010observation,19-yang2017anti,20-zhang2019dynamically}. In particular, a reduction in the dimensionality of the eigenmode boosts the sensor sensitivity to external perturbations, and the reaction of EP-based sensor to a perturbation ($\varepsilon\ll1$) is proportional to  $\sqrt[N]{\varepsilon}(N\in\mathbb{R})$, where N is the order of the EP \cite{21-wiersig2014enhancing,22-wiersig2016sensors,23-hodaei2017enhanced}. 

During the past few years, non-Hermitian optical gyroscopes operated at second-order EPs have been investigated in various photonic structures. The response of PT-symmetric ring laser gyroscopes (RLGs) and anti-PT-symmetric gyroscopes operating at a second-order EP exhibits a square-root dependence on rotational rates, resulting in frequency splitting several orders of magnitude higher than those of traditional gyroscopes\cite{08-ren2017ultrasensitive,10-de2019high,11-wang2024novel}. M. Khajavikhan et al. and Lai et al. demonstrated experimental implementations of the non-Hermitian RLGs , achieving predicted EP-enhanced scale factors compared to conventional gyroscopes\cite{05-hokmabadi2019non,06-lai2019observation}. To further improve the robustness of the non-Hermitian gyroscope, Li et al. proposed a scheme that utilizes an exceptional surface constructed of numerous second-order chiral EPs to enhance rotation sensing, achieving an improvement of nearly four orders of magnitude in sensitivity\cite{24-li2021exceptional}. 

Moreover, HOEPs offer substantial potential for enhancing light-matter interactions. Given that $\sqrt[N]{\varepsilon }$, this feature opens up new possibilities for designing ultrasensitive gyroscopes based on such Nth-order EPs \cite{07-jiang2019chip,25-zhong2020hierarchical}. However, operating the non-Hermitian photonic systems at HOEPs to achieve extreme response requires strict precise controlling parameters; hence, the implementation of HOEPs in practical photonic schemes is exceptionally challenging \cite{23-hodaei2017enhanced,26-xie2024enhanced}.

In this Letter, we present a novel construction approach for generating HOEP in anti-parity time symmetric photonic configuration. This strategy enables the construction of a fourth-order EP via two indirectly coupled anti-parity time symmetric cavities, significantly reducing the parameter requirements for tuning the system to HOEP. We theoretically demonstrate enhanced frequency splitting and noise-limited precision improvements compared to gyroscopes based on second-order EP. This approach can be leveraged to integrate robustness with improved sensitivity in HOEP-based non-Hermitian gyroscopes.

\section{Non-Hermitian HOEP System for Enhanced Rotation Detection}
\begin{figure*}[ht]
\centering
\includegraphics[width=\linewidth]{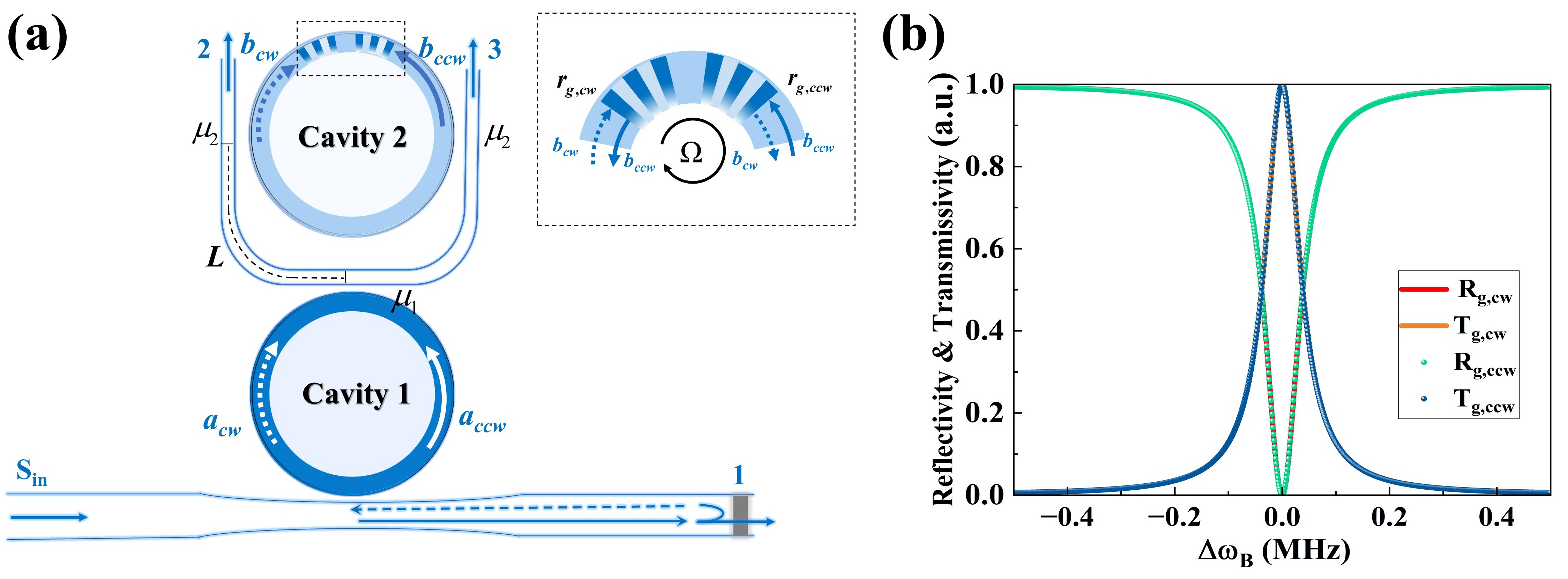}
\caption{(a) Schematic diagram of the fourth-order EP anti-PT symmetric photonics structure.  Particularly, a $\pi$-phase shifted Bragg grating is inserted into cavity2. The black (dashed line) box demonstrates the reflection between the CW and CCW modes at the grating. The intrinsic loss of the cavity1(cavity2) is $\gamma_{c1}(\gamma_{c2})$, $\mu_1(\kappa_{ex})$ is the external coupling between the up (down) waveguide of the cavity1, and $\mu_2$ is the external coupling coefficient of the cavity2. (b) Reflection spectra and transmission spectra of a $\pi$-PBG at $\Omega=0$ with an effective index $n_{eff}=1.44$, a uniform index modulation $\Delta n_{ac}=5\times10^{-4}$, a Bragg wavelength $\lambda_B=1550nm$ and length $l=8mm$. }
\label{fig1}
\end{figure*}
A schematic of our configuration is depicted in Fig.\ref{fig1}(a), where two whispering-gallery-mode (WGM) cavities, each supporting a pair of counterpropagating modes, are indirectly coupled through a waveguide with an indirect coupling coefficient $\kappa=\sqrt{\mu_1\mu_2}e^{-i\theta}$, where the transmission phase in the waveguide is $\theta=2\pi n_g L/\lambda$, $n_g$ is the group index and $\lambda$ is the operating wavelength. To realize the anti-PT symmetric criteria\cite{19-yang2017anti}, an appropriate distance L between the two cavities should be controlled to implement $\theta=(2m+1)\frac{\pi}{2}(m\in\mathbb{N})$ in the experiment, achieving an imaginary coupling condition, i.e., $\kappa =-i \sqrt{\mu_1\mu_2}$  with $\theta=\pi/2$. A partial reflector with reflectivity $r_m$ is on one side of the waveguide, causing the unidirectional coupling from $a_{ccw}$ to $a_{cw}$ with unidirectional coupling strength $\kappa_0=\kappa_{ex} r_m e^{2i\phi}$, where $\phi$ is the propagation phase in the straight waveguide. In addition, our previous work \cite{24-li2021exceptional} proposed a mechanism that relates the rotation to the additional mode coupling that utilized a $\pi$-phase shifted Bragg grating ($\pi$-PBG). To enhance the rotation detection, a $\pi$-PBG with reflectivity $r_g(\omega_b)$ was inserted into the cavity2, introducing the mode coupling between $b_{cw}$ and $b_{ccw}$ as shown in the inset of Fig.1(a). Within the rotating-wave frame, the coupled mode equation can be described as
\begin{equation}
\begin{array}{l}
i\frac{{d{a_{ccw}}}}{{dt}} = ({\omega _a} - i{\gamma _1}){a_{ccw}} + \kappa  {b_{cw}} - i\sqrt {{\kappa _{ex}}} {s_{in}},\\
i\frac{{d{b_{cw}}}}{{dt}} = ({\omega _b} - i{\gamma _2}){b_{cw}} + \kappa {a_{ccw}} + \varepsilon {b_{ccw}},\\
i\frac{{d{a_{cw}}}}{{dt}} = ({\omega _a} - i{\gamma _1}){a_{cw}} + \kappa {b_{ccw}} + {\kappa _0}{a_{ccw}} - i{r_m}{e^{2i\phi }}\sqrt {{\kappa _{ex}}} {s_{in}},\\
i\frac{{d{b_{ccw}}}}{{dt}} = ({\omega _b} - i{\gamma _2}){b_{ccw}} +\kappa  {a_{cw}} + \varepsilon {b_{cw}},
\end{array}
\label{EQ1}
\end{equation}
here, $\omega_a$ and $\omega_b$ are the resonance frequency of the cavity1 and cavity2 respectively,  ${\gamma_1}=(\gamma_{c1} + \mu_1 + \kappa_{ex})/2$ and ${\gamma_2}=(\gamma_{c2} + 2\mu_2)/2$ are the effective loss rate of cavity1 and cavity2, respectively. The coupling strength between the $b_{cw}$ and the $b_{ccw}$ denotes $\varepsilon=\frac{r_g(\omega_b) c}{2\pi n_g R_2}$ with the radius $R_2$ of the cavity2\cite{z510.1063/5.0171249,8955810,201500207}. With the coupled-mode theory,  this feature can be well described by a non-Hermitian Hamiltonian  
\begin{equation}
H_{\text{eff}} =
\begin{bmatrix}
\omega_a - i\gamma_1 & \kappa & 0 & 0 \\
\kappa & \omega_b - i\gamma_2 & 0 & \varepsilon \\
\kappa_0 & 0 & \omega_a - i\gamma_1 & \kappa \\
0 & \varepsilon & \kappa & \omega_b - i\gamma_2
\end{bmatrix}.
\label{EQ2}
\end{equation}
Based on the Eq.(\ref{EQ2}), it is necessary to make  $\varepsilon = 0$ under non-rotating conditions to operate the system at EP. To satisfy this requirement, we need $r_g(\omega_b) = 0$. As depicted in Fig.~1(a), when the system rotates in the CW direction with rotation rate $\Omega$, $b_{cw}$ and $b_{ccw}$ modes experience an opposite Sagnac frequency shift $\Delta\omega_{s2}=4\pi R_2/n_g \lambda$, where $\lambda$ is the operating wavelength. The field reflection coefficients from $b_{cw}$ ($b_{ccw}$) to the $b_{ccw}$ ($b_{cw}$) is $r_{g,cw}\left( \omega_b - \Delta \omega_{s2} \right)$ ($r_{g,ccw}\left(\omega_b + \Delta \omega_{s2} \right)$). Hence, the field reflectivity $r_{g,cw/ccw}$ and transmissivity $t_{g,cw/ccw}$ are related to the Sagnac shift $\Delta\omega_{s2}$,i.e.,
\begin{equation}
 {r_{g,cw/ccw}} = \left| { - \frac{{2{\delta _{cw,ccw}}r_1^2{\kappa _g}}}{{{\delta _{cw,ccw}}^2(1 + r_1^2) - 2i{\delta _{cw,ccw}}{\sigma _{cw,ccw}}{r_1} - \kappa _g^2(1 - r_1^2)}}} \right|
\label{EQ3}
\end{equation}
and $t_{g,cw/ccw}=\sqrt{1-{r_{g,cw/ccw}^2}}$, where the detuning parameter is $\delta_{cw,ccw} = n_{eff}(\Delta \omega_B \pm \Delta \omega_{s2})/c$, $\Delta \omega_B = \omega_b-\omega_B$ with the Bragg frequency $\omega_B$, $\sigma_{cw,ccw} = \sqrt{\kappa_g^2 - \delta_{cw,ccw}^2}$, $\kappa_g = \Delta n_{ac} / \lambda$ is the coupling coefficient of the grating, $\Delta n_{ac}$ is the refractive modulation strength, $r_1 = \tanh (\kappa_g l)$, and $l$ is the length of the grating\cite{24-li2021exceptional,z4PhysRevA.102.013515}. The reflectivity and transmissivity of power are $R_{g,cw/ccw} = |r_{g,cw/ccw}|^2$ and $T_{g,cw/ccw} = |t_{g,cw/ccw}|^2$  at   $\Omega = 0$ are illustrated in Fig.\ref{fig1}(b) . A distinct narrow dip around $\Delta \omega_B = 0$ is observed in the reflection spectrum of the $\pi$-PBG, which is essential for the operating system at EP. Specifically, when the resonance frequency $\omega_b$ equals $\omega_B$, the reflectivity is close to zero, corresponding to $\varepsilon=0$. 

Now, the Eq.(\ref{EQ2}) is reformulated into a new expression defined as $H_0 = \begin{bmatrix}H_a&\hat{0} \\ \hat{\kappa_0}&H_a\end{bmatrix}$, where $H_a = \begin{bmatrix}\omega_a-i\gamma_1&\kappa \\ \kappa&\omega_b-i\gamma_2\end{bmatrix}$ , $\hat{\kappa_0}$ is a $2\times 2$ matrix and $\hat{0}$ is a null matrix. Considering the parameters required to realize an EP in the anti-parity symmetric system \cite{19-yang2017anti}, i.e., $\gamma_1=\gamma_2=\gamma$, $\Delta_a = -\Delta_b=\Delta$, $\Delta_{a,b}=\omega_{a,b}-\omega_0$ with $\Delta = (\omega_a - \omega_b) / 2$ and  $\omega_0 = (\omega_a+\omega_b) / 2$,  the eigenfrequencies of $H_0$ are $\omega_{EP,n}=\omega_0-i\gamma\pm\sqrt{\Delta^2+\kappa^2}$, (n=1,2,3,4). When $|\Delta| = \sqrt{\mu_1 \mu_2}$, the eigenfrequencies of $H_0$ are degenerate at $\omega_{EP} = \omega_0 - i\gamma$ and the corresponding eigenvectors coalesce to one eigenvector $\tilde{\alpha}_{EP} = [0, 0, i, 1]^T$, suggesting the generation of the fourth-order EP in the proposed system. The formula of $H_0$ is similar to the Hamiltonian of exceptional surface systems\cite{z1-Zhong:19,z2,25-zhong2020hierarchical,z510.1063/5.0171249},   significantly, the physical parameter $\kappa_0$ does not contribute to the construction of the EP, indicating that the physical parameters required to construct a fourth-order EP are identical to those needed for a second-order EP. Compared to the general approach of constructing Nth-order EPs using N-coupled cavities, this scheme can eliminate the complexity of multi-parameter tuning and potentially promote the robustness of anti-PT symmetric systems with HOEPs\cite{z3Wu:21}.

\begin{figure*}[ht]
\centering
\includegraphics[width=\linewidth]{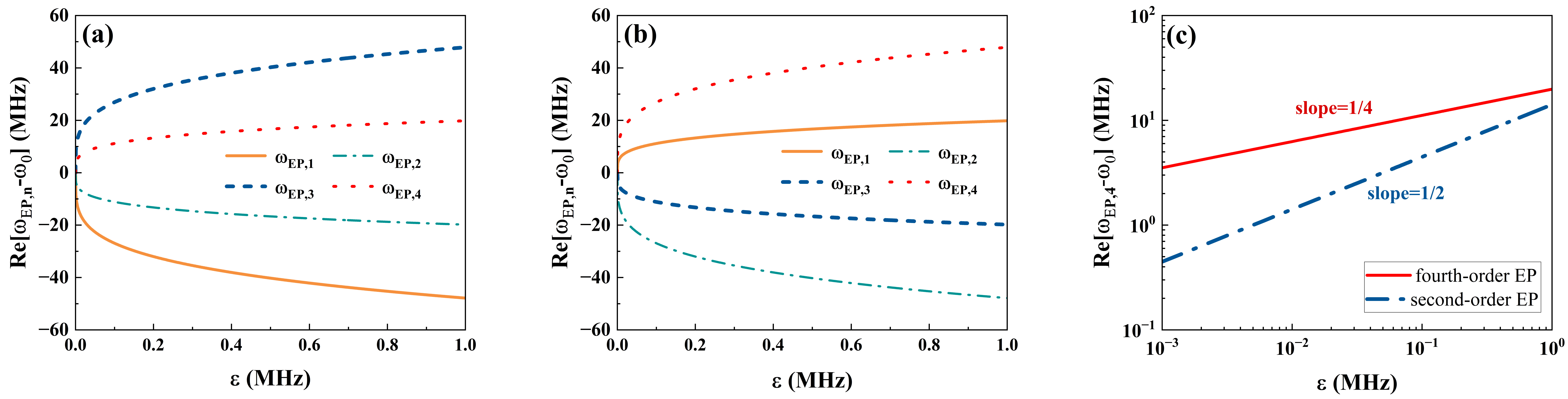}
\caption{(a) Real and (b) imaginary splitting of the eigenfrequency at fourth-order EP as a function of a perturbation $\varepsilon$.  (c) Logarithmic scale plot of the real eigenfrequency  splitting $\text{Re}[\omega_{EP,4} - \omega_0]$ with the variation of $\varepsilon$. Here, $\phi=\pi/4$.}
\label{fig2}
\end{figure*}

When the system rotates, the perturbed Hamiltonian is given by
\begin{equation}
H_{I} = \begin{bmatrix}
-\Delta \omega_{s1} & 0 & 0 & 0 \\
0 & \Delta \omega_{s2}  & 0 & \epsilon(-\Delta \omega_{s2}) \\
0 & 0 & \Delta \omega_{s1} & 0 \\
0 & \epsilon(\Delta \omega_{s2}) & 0 & -\Delta \omega_{s2} 
\end{bmatrix}.
\label{EQ4}
\end{equation}
Assuming the rotation rate $\Omega$ is sufficiently small, i.e., $\Delta {\omega _{s1,2}} \ll \varepsilon (\Delta {\omega _{s2}})$, the eigenfrequency of the $H=H_0+H_I-\omega_0I$ is
\begin{equation}
\omega_{EP,n}-\omega_0=i^n e^{\frac{i\pi}{4}}|\Delta|^{1/2}\kappa_0^{1/4}\varepsilon(\Delta\omega_{s2})^{1/4}+\mathcal{O}(\varepsilon(\Delta\omega_{s2})),
\label{EQ5}
\end{equation}
where n=1,2,3,4. According to Eq.(\ref{EQ5}), the eigenfrequency diverges with the fourth root of $\varepsilon(\Delta\omega_{s2})$ and is fundamentally governed by the parameters $\Delta$ and $\kappa_0$. To evaluate the sensitivity of this system, the complex eigenfrequency splitting can be calculated as 
\begin{equation}
\Delta {\omega _{EP}} \approx 2e^{\frac{i\pi}{4}}|\Delta|^{1/2}\kappa_0^{1/4}\varepsilon {(\Delta {\omega _{s2}})^{\frac{1}{4}}},
\label{EQ6}
\end{equation}
 The real part of $\Delta\omega_{EP}$ represents the frequency splitting, while the imaginary part corresponds to linewidth broadening. Fig.\ref{fig2}(a) and Fig.\ref{fig2}(b) depict the real parts and imaginary parts of the four eigenfrequencies with the variation of the perturbation strength $\varepsilon$, respectively. As predicted, the perturbation lifts the degeneracy of the eigenfrequencies, causing the eigenstate to split into four different eigenmodes. Fig \ref{fig2}(c) presents the real part of the eigenfrequency $\omega_{EP,4}-\omega_0$ as a function of $\varepsilon$ on a logarithmic scale, notably, the slope is 1/4, indicating the fourth-root dependence predicted by Eq.(5).

\section{NUMERICAL RESULTS}
\subsection{Enhancement of eigenfrequency splitting and Scale factor in non-Hermitian fourth-order EP system }

\begin{figure*}[ht]
    \centering
    \includegraphics[width=\linewidth]{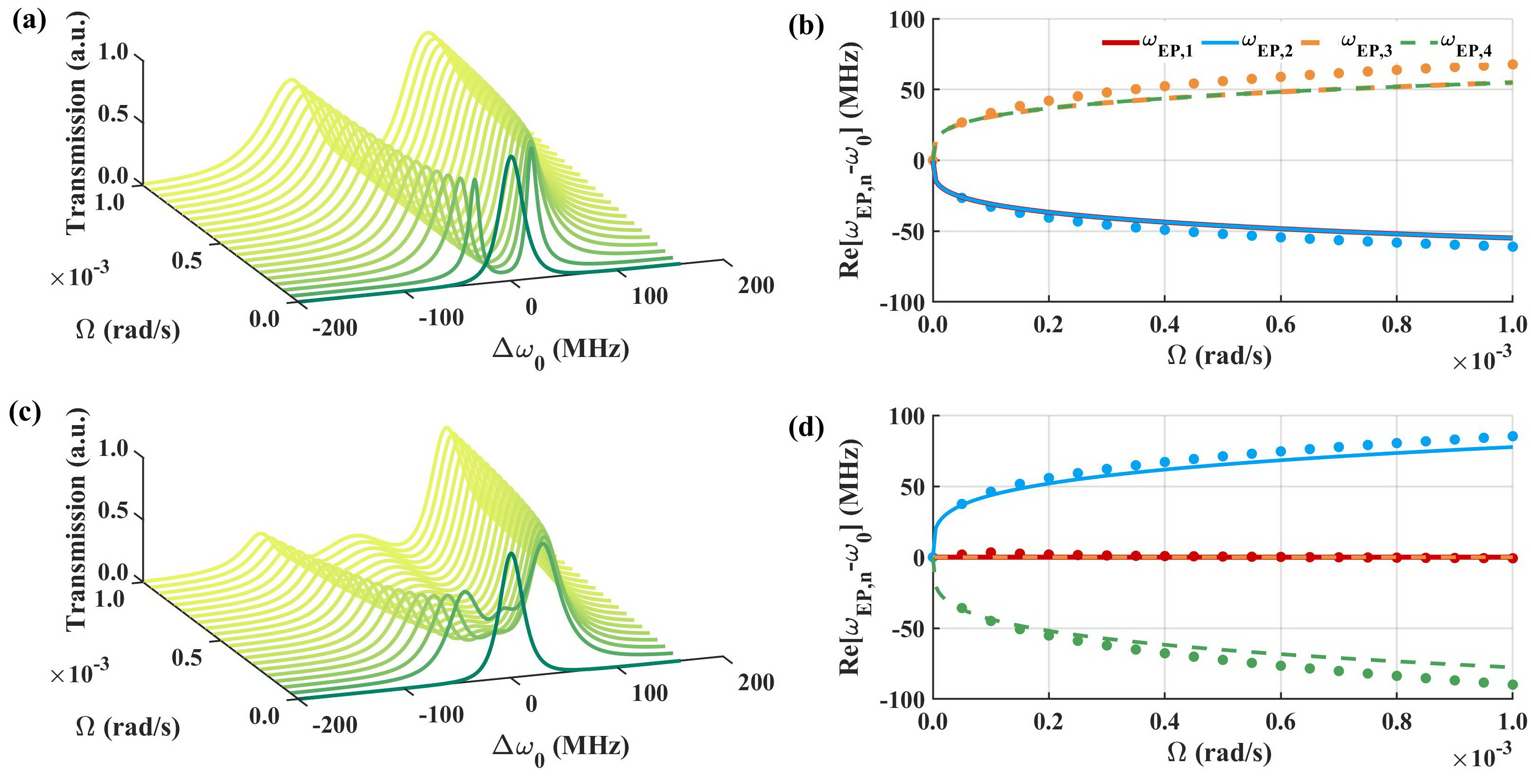}
    \caption{ Transmission spectrum of the fourth-order EP system at (a) $\phi=0$  and (c) $\phi=\pi/2$. The scatters in (b) and (d) correspond to the peak trajectories of the transmission spectrum in (a) and (c), respectively. The color lines emphasize the  eigenfrequencies [given in Eq.(\ref{EQ5})] with the variation of rotation rate. The simulated parameters are $\gamma_{c1}=\gamma_{c2}=2\times10^8 Hz$, $\kappa_{ex}=2\times10^8Hz$, $\mu_1=\mu_2=2\times10^8Hz$, $\gamma_1=\gamma_2=2\times10^7Hz$, $R_2= 4.95mm$ and $r_m = 0.9$.}
    \label{fig3}
\end{figure*}

The rotation-induced eigenfrequency splitting can be extracted from the transmission spectrum. Assuming that the system is probed via the waveguide in port2, based on the input-output relation $S_{out}=\sqrt{\mu_2}b_{cw}+\sqrt{\mu_1}a_{ccw}e^{i\theta}$, the transmission spectrum can be obtained by solving Eq.~(1) in the steady-state approximation
\begin{equation}
\begin{aligned}
T &= \left| \frac{\chi_1 \sqrt{\kappa_{ex} \mu_1} e^{i \theta} + \chi_2 \sqrt{\kappa_{ex} \mu_2}}
{(\mathcal{R}_1^2 (\mathcal{R}_2^2+\varepsilon^2)+2\mathcal{R}_1\mathcal{R}_2\kappa^2+\kappa^4-\varepsilon\kappa^2\kappa_0} \right|^2, \\
\chi_1 &= \mathcal{R}_1(\mathcal{R}_2^2+\varepsilon^2) +(\mathcal{R}_2 - ir_m\varepsilon)\kappa^2, \\
\chi_2 &=i\kappa(\mathcal{R}_1 (\mathcal{R}_2 +ir_m \varepsilon)+\kappa^2-\varepsilon \kappa_0),
\end{aligned}
\label{EQ7}
\end{equation}

where $\mathcal{R}_{1}=-i\Delta\omega_{a}+\gamma_{1}$, $\mathcal{R}_2=-i\Delta\omega_b+\gamma_2$ and $\Delta\omega_{a,b}=\omega_{a,b}-\omega$ with incident laser frequency $\omega$. Considering the unidirectional coupling coefficient $\kappa_0 = \kappa_{ex} r_m e^{2i\phi}$, it indicates that the propagation phase $\phi$ will provide an additional degree of freedom to control the complex eigenfrequency splitting. As a result, by adjusting the phase, rotation can produce a distinct change in the real component of eigenfrequencies, thereby enabling the observation of distinct spectral responses. Fig.\ref{fig3}(a) and Fig.\ref{fig3}(c) plot a series of transmission spectra as functions of rotation rate for phases $\phi=0$ and $\phi=\pi/2$, respectively. From Fig.\ref{fig3}(b) and Fig.\ref{fig3}(d), one can see the peak trajectories of the transmission spectrum aligned with the real part of the eigenfrequencies as predicted by Equation (\ref{EQ5}), which means that the eigenfrequencies splitting can be extracted from the transmission spectrum. Eigenfrequencies splitting is anticipated to be detectable at lower rotation rates, when the gain introduced into the resonator compensates for the eigenmode linewidth. 

\begin{figure}[ht]
    \centering
    \includegraphics[width=0.65\linewidth]{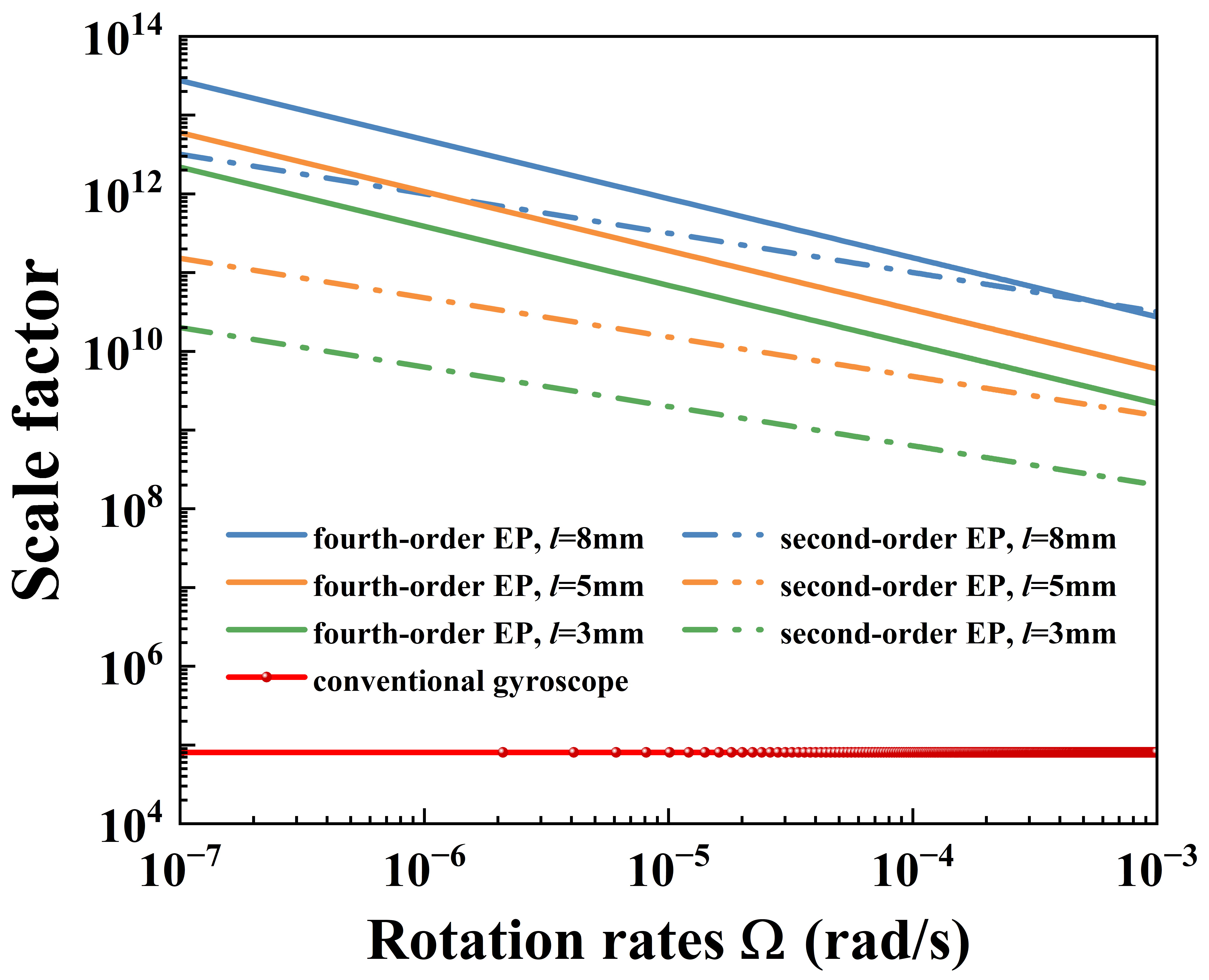}
    \caption{ Comparison of the Sagnac scale factor between EP-based gyroscope and conventional gyroscope within $\phi=0$. }
    \label{fig4}
\end{figure}

In general, the scale factor of a gyroscope is a critical parameter, typically defined by the derivative of the eigenfrequency splitting with respect to the rotation rate  $\Omega$. The Sagnac scale factor for the proposed fourth-order EP system is
\begin{equation}
    S_f= {\left. {\frac{{\partial {\Delta\omega _{EP}}}}{{\partial \Omega }}} \right|_{\Omega  \to 0}} = \frac{1}{2}{e^{\frac{i\pi}{4}}}|\Delta|^{1/2}\kappa_0^{1/4}{\varepsilon ^{ - 3/4}}\frac{c}{{{n_g}2\pi {R_2}}}\frac{{\partial {r_{g}}}}{{\partial (\Delta {\omega _{S2}})}}\frac{{4\pi {R_2}}}{{{n_g}\lambda }}.
 \label{EQ8}
\end{equation}
The $4\pi R/n_g\lambda$ arising from the Sagnac effect is the scale factor of the conventional gyroscope and  ${S_e} = \frac{1}{2}{e^{\frac{i\pi}{4}}}|\Delta|^{1/2}\kappa_0^{1/4}{\varepsilon ^{-3/4}}\frac{c}{{{n_g}2\pi {R_2}}}\frac{{\partial {r_{g}}}}{{\partial (\Delta {\omega _{S2}})}}$is the fourth-order EP enhancement factor. This enhancement factor depends on the $\Delta$, $\kappa_0$ and the reflection amplitude $r_g$ determined by the structure parameters of  $\pi$-FBG. As depicted in Fig.\ref{fig4}, the scale factor increases with  increasing of  grating length, this is due to the greater perturbation induced by rotation with longer grating length. The superior performance of the fourth-order EP-based gyroscope becomes prominent at low rotation rates. When the grating length is $l=5mm$, the scale factor of the gyroscope in the vicinity of the fourth-order EP is enhanced by two orders of magnitude compared to a gyroscope operating at the second-order EP, and by nearly eight orders of magnitude compared to a conventional gyroscope. 

\subsection{Noise limits in non-Hermitian fourth-order EP system}
Although the EP sensor demonstrates an enhancement in the eigenfrequency splitting, this does not result in improved detection precision under noise-limited conditions, as the non-orthogonality of modes induced by EP increases frequency noise due to the enhanced Petermann factor\cite{30-lau2018fundamental,wang2020petermann}. In fact, for a conventional
gyroscope, the Sagnac frequency shift can be measured as a change in the transmission spectrum for a given input power $P_{in}$. When the gyroscope rotates with a rotation rate $\Omega$, the transmission coefficient experiences a change in $\Delta T$, resulting in a change in the transmission signal $P_{out}=P_{in}\Delta T$ . The sensitivity can be defined as follows
\begin{equation}
    S = {\left. {\frac{1}{{{P_{{\rm{in}}}}}}\left({\frac{{d{P_{{\rm{out}}}}}}{{d\Omega }}} \right)} \right|_{\Omega  \to 0}} = {\left. {\frac{{dT}}{{d\Omega }}} \right|_{\Omega  \to 0}}.
    \label{EQ9}
\end{equation}
Another critical physical quantity evaluating gyroscope performance is the angular random walk (ARW) defined as $ARW=\sigma_s/S P_{in}$, which quantifies the minimum detectable rotation rate under noise limitations. The total noise $\sigma_s$  originates from shot noise, detector noise, laser relative intensity noise, laser frequency noise, and other sources. Utilizing balanced detection techniques and enhancing the Q factor can effectively minimize the noise sources, predominantly limiting them to shot noise and detector noise, and then the ARW is
\begin{equation}
    ARW = \frac{{\sqrt {2(\hbar \omega  + q/\rho ){P_{{\rm{out}}}} + \sigma _{{\rm{DN}}}^2} }}{{S{P_{in}}}},
 \label{EQ10}
\end{equation}
here,  $\sigma _{shot}^2 = 2(\hbar\omega  + q/\rho ){P_{{\rm{out}}}}$ is the shot noise, $\sigma _{DN}^2$ denotes the detector noise, $\hbar\omega$ is the photon energy, q is the electron charge, and $\rho$ denotes the detector responsivity. 

\begin{figure}[ht]
    \centering
    \includegraphics[width=0.65\linewidth]{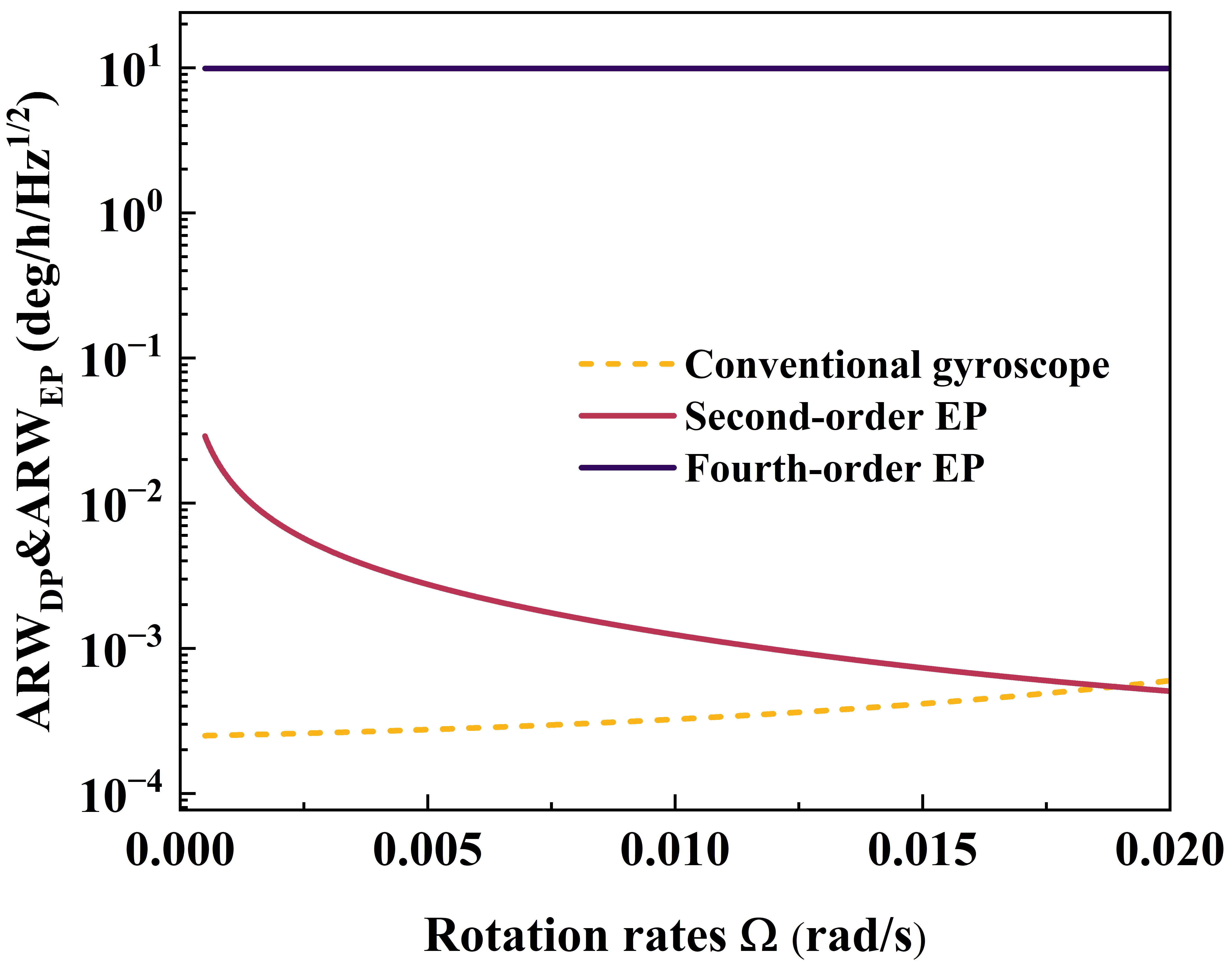}
    \caption{ The ARW of the EP-based gyroscope and conventional gyroscope plotted versus rotation rates for $l=8mm$ and $\phi=0$.}
    \label{fig5}
\end{figure}

 Fig.\ref{fig5} compares the ARW of the EP-based gyroscope and the optimized ARW of the conventional gyroscope. According to the contrast principle of the gyroscopes, both the EP gyroscope and a conventional single-resonator gyroscope with the same perimeter and intrinsic loss rate are considered. The simulated parameters are set as  radius $R_2$=4.95mm of cavity2, input laser with wavelength $\lambda$= 1550nm and power $P_{in}=100\mu W$, responsivity $\rho=1A/W$, and detector noise $\sigma_{DN}^2=0.4pW/\sqrt{Hz}$. The gyroscope operating at an EP is shown to exhibit superior precision compared to conventional gyroscopes, and the reasons for this improved detection precision are as follows.  For the HOEP-based non-Hermitian gyroscopes presented here, rotation introduces a larger perturbation compared to a conventional gyroscope. As described in Eq. (4),  rotation affects the field transmission coefficient through two mechanisms: (i) the Sagnac frequency shift $\Delta\omega_s$ and (ii) the additional coupling $\epsilon$ induced by rotation. Furthermore, at a second-order EP or a fourth-order EP, the presence of two or four collapsing eigenmodes results in a single transmission peak. When rotation breaks the degeneracy of these eigenmodes, they separate, resulting in a reduced interaction. Consequently, the transmission experiences a sharp change around the EP to generate enhanced sensitivity. 

\begin{figure}[ht]
    \centering
    \includegraphics[width=\linewidth]{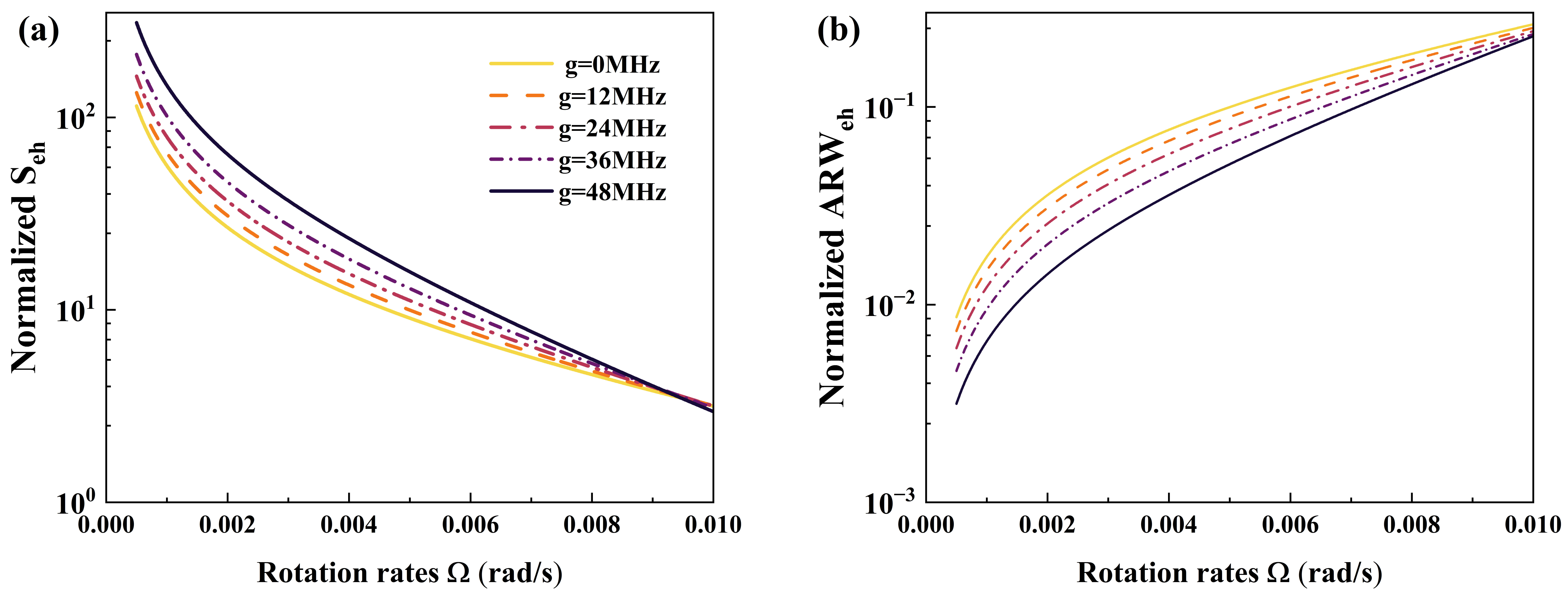}
    \caption{(a) Sensitivity enhancement factor $S_{eh}$ and (b) Normalized $ARW_{eh}$ factor versus rotation rates for $l=8mm$ and $\phi=0$.}
    \label{fig6}
\end{figure}

Additionally, to assess the ARW of a gyroscope at fourth-order EP compared to that at second-order EP, we introduce the sensitivity enhancement factor $S_{eh}=S_{EP,4}/S_{EP,2}$ and normalized ARW factor $ARW_{eh}=ARW_{EP,4}/ARW_{EP,2}$, here, the subscripts 4 and 2 represent the condition of fourth-order EP and second-order EP respectively. Fig.\ref{fig6}(a) and Fig.\ref{fig6}(b) illustrate the sensitivity enhancement and the normalized ARW factor for different gain coefficients (g), respectively. To ensure system stability, the effective loss rate $\gamma=\gamma_{1,2}-g$ must be positive. The ARW of the gyroscope operating at fourth-order EP is reduced  by two orders of magnitude compared to that operating at a second-order EP.  When the gain is introduced, the sensitivity enhancement factor increases and the corresponding normalized ARW factor decreases, indicating improved measurement precision.  It is evident that the normalized ARW factor is reduced, as indicated by the variations in transmittance observed at higher gain coefficients.

\section{Conclusions}
In conclusion, we have presented a framework for a more robust implementation of an anti-parity time symmetric gyroscope with HOEPs. A fourth-order EP is achieved via two indirectly coupled microresonators, reducing the complexity of parameters required for building HOEPs. We have demonstrated an enhancement in frequency splitting and a reduced ARW compared to the anti-PT symmetric gyroscope operating at a second-order EP. We showed that the transmittance spectrum varies significantly at higher gain compensation coefficients, further enhancing the detection precision of the HOEP-based gyroscope. The proposed configuration is expected to inspire the development of integrated photonic gyroscopes with HOEPs. 

\begin{backmatter}
\bmsection{Funding}
National Key Research and Development Program of China (2021YFB3900701); National Natural Science Foundation of China (61975005, 12104252).

\bmsection{Disclosures}
The authors have no conflicts of interest to disclose.

\medskip

\bmsection{Data Availability Statement}
The data that support the findings of this study are available from the corresponding author upon reasonable request.

\end{backmatter}

\bibliography{Optica-template}

\end{document}